\documentclass[prd,floatfix,twocolumn]{revtex4}
\usepackage{graphicx,ulem}
\usepackage{graphics,epsfig}
\usepackage{amssymb}
\newcommand{\ETAL}{{et al.}}
\newcommand{\be}{\begin{equation}}
\newcommand{\ee}{\end{equation}}
\newcommand{\ba}{\begin{eqnarray}}
\newcommand{\ea}{\end{eqnarray}}
\newcommand{\bc}{}
 
\newcommand{\fns}{\footnotesize}

\newcommand{\fnsc}{\scriptsize}
\begin{document}
\title{Constraining Isocurvature Initial Conditions with WMAP 3-year data}

\author{Rachel Bean$^\sharp$, Joanna Dunkley$^{\dagger, \flat}$, Elena Pierpaoli
$^{\ddagger}$}
\affiliation{$^\sharp$ Dept. of Astronomy, Cornell University, Ithaca, NY, USA.\\
$^\dagger$ Dept. of Physics, Princeton University, Princeton, NJ, USA. \\$^\flat$ Dept. of Astrophysical Sciences, Princeton University, 
Princeton, NJ, USA.\\$^\ddagger$ 
Physics and Astronomy Department, University of Southern California, Los Angeles, CA 90089-0484, USA}

\begin{abstract}
%
We present constraints on the presence of isocurvature modes from the temperature and polarization CMB spectrum 
data from the WMAP satellite alone, and in combination with other datasets including SDSS galaxy survey and SNLS supernovae.
We find that the inclusion of polarization data allows the WMAP data alone, as well as in combination with complementary observations, to place improved limits on the contribution of CDM and neutrino density isocurvature components individually. 
With general correlations, the upper limits on these sub-dominant isocurvature components are reduced to $\sim$60\% of the first year WMAP results, with specific limits depending on the type of fluctuations.  If multiple isocurvature components are allowed, however, we find that the data still allow a majority of the initial power to come from isocurvature modes. As well as providing general constraints we also consider their interpretation in light of specific theoretical models like the curvaton and double inflation. 
\end{abstract}
\maketitle
\section{Introduction}\label{intro}

Impressive recent developments in measurements of the Cosmic Microwave 
Background  (CMB)   temperature and polarization anisotropies 
\cite{Hinshaw:2006ia,Jarosik:2006ib,Page:2006hz,Spergel:2006hy}, large 
scale structure (LSS) 
\cite{2df,Tegmark:2003uf,Seljak:2004sj,Eisenstein:2005su,Cole:2005sx},
supernovae \cite{Riess:2004nr,Astier:2005qq} and the Lyman-$\alpha$
 forest
  \cite{Seljak:2006} have enabled stringent testing 
of the cosmological model including the composition of the universe and the 
initial conditions that seeded inhomogeneities. Although the simplest initial 
conditions, arising from single field inflation, predict scale-invariant, 
adiabatic inhomogeneities, this is far from the only possibility. Isocurvature 
modes are predicted by a wide range of scenarios, for example multi-field 
inflation \cite{Polarski:1994rz,Garcia-Bellido:1995qq,Linde:1996gt,Pierpaoli:1999zj,Kawasaki:2001in}, topological defects \cite{Battye:1998xe,Battye:1999em}, 
and through the decay of particles prior to nucleosynthesis such as a scalar 
curvaton \cite{Lyth:2001nq,Moroi:2001ct,Bartolo:2002vf,Moroi:2002rd,Lyth:2002my,Dimopoulos:2003ii,Dimopoulos:2003az}, axions \cite{Bozza:2002fp} or the 
Affleck-Dine model of baryogenesis \cite{Kawasaki:2001in}. The degree of 
correlation with adiabatic perturbations can vary across the full spectrum, 
from completely (anti) correlated in the curvaton scenario, intermediate 
correlation in some multi-field inflation models, to 
uncorrelated modes in cosmic string scenarios.
Most proposed scenarios generate solely
baryon or Cold Dark Matter (CDM) isocurvature modes \cite{cdm_theory} 
; mechanisms generating neutrino density and velocity isocurvature modes 
are also possible \cite{nu_theory,BMT:2000}, though the latter are more difficult to motivate.
 
Although the simplest adiabatic scenario is in complete agreement with 
the current data \cite{Spergel:2006hy}, the question still arises, how large a 
contribution could isocurvature modes make and how sensitive is the 
cosmological parameter estimation to their inclusion?  In this paper we 
confront these questions, specifically focusing on constraints from the 
WMAP CMB temperature and polarization power spectra 
\cite{Jarosik:2006ib,Hinshaw:2006ia, Page:2006hz,Spergel:2006hy},  SDSS galaxy 
matter power spectrum \cite{Tegmark:2003uf} and the SNLS supernovae survey 
\cite{Astier:2005qq}. Experiments in the past ten years have ruled out models 
with purely isocurvature perturbations 
\cite{Stompor:1995py,Langlois:2000ar,Enqvist:2000hp,Amendola:2001ni}, but 
have not excluded those with admixtures of adiabatic and isocurvature modes.  
Constraints placed on models with a single 
isocurvature mode, in light of the 
first year WMAP results, indicate that it must be sub-dominant 
\cite{Peiris:2003ff, Valiviita:2003ty,crotty:03,Gordon:2003hw,beltran:04,Moodley:2004nz,Kurki-Suonio:2004mn,Beltran:2005gr},  
but models with multiple modes may have larger non-adiabatic contributions 
\cite{Bucher:2004an, Moodley:2004nz,Dunkley:2005va}. Here we analyze the 
constraints in light of the 3-year WMAP data release of temperature and 
polarization data for which only an analysis for perfectly correlated 
CDM  or baryon isocurvature perturbations has been conducted so far 
\cite{Lewis:2006ma,Seljak:2006}.

In section \ref{method} we outline the approach and parameterizations used in the analysis. In sections \ref{puresingiso} and \ref{gensingiso} we respectively establish the constraints on scenarios with purely correlated and generally correlated single isocurvature modes. In doing so we reflect on what the 
limits on isocurvature contributions imply for some key particle 
based theories which depart from purely adiabatic perturbations. In section \ref{mixed} we expand our analysis to allow multiple isocurvature modes with general correlations, in which destructive cancellations can allow a large fractional isocurvature contribution. We conclude with a summary of our findings and implications for future cosmological observations in section \ref{Conclusion}.

\section{Approach}\label{method}

Perturbations to the metric may give rise to both 
curvature perturbations on comoving 
hypersurfaces, as well as entropy perturbations where the 
space-time curvature vanishes at early times. The former are 
termed 
adiabatic perturbations and may be quantified by the curvature perturbation, $\cal R$.
The latter are isocurvature modes, quantified by the entropy perturbation 
${\cal S}_x = \delta \rho_x/(\rho_x+p_x) - \delta\rho_\gamma/(\rho_\gamma+p_\gamma)$ 
in the case of density perturbations, $\delta\rho$, between photons and a fluid $x$, 
 which may be CDM or baryons. There are two 
further isocurvature modes 
where the sum of the neutrino and photon densities, or momentum densities, 
are initially unperturbed, whose initial conditions are given in \cite{BMT:2000,camb}.

To first order, initial conditions $\delta\rho_c = -\delta\rho_b $, $\delta\rho_{\gamma}=\delta\rho_{\nu} = 0$ allow a time independent solution to the perturbation equations for the pressureless matter components. Baryon and CDM isocurvature initial conditions, up to a factor of $\Omega_c/\Omega_b$, are essentially observationally indistinguishable
and therefore we only consider CDM isocurvature scenarios of the two here.

The perturbations may be characterized using the covariance matrix 
$\Delta$, where
\ba
\Delta_{ij}(k)\delta^3({\bf k}-{\bf k'}) = 
\left(\frac{k}{2\pi} \right)^3\langle{ \chi}_i({\bf k})\,{ \chi}_{j}^*({\bf k'})\rangle
\ea
as shown in \cite{BMT:2000}. 
The subscript $i$ may take the values \{A,C,N,V\}, 
labeling the adiabatic (AD), CDM isocurvature (CI), neutrino 
isocurvature density (NID), and neutrino isocurvature velocity (NIV) 
components respectively, 
and the random variable $\chi_i$ corresponds to the amplitude of the $i$th mode.

We assume that the elements of the matrix $\Delta$ can be parameterized as
\ba
\Delta_{ij}(k) = {\cal A}_{ij}~\left(\frac{k}{k_0}\right)^{n_{ij}-1}, \qquad n_{ij} =
{1 \over 2}(n_i+n_j)
\ea
in terms of a set of amplitude parameters ${\cal A}_{ij}$, power-law 
spectral indices $n_i$, and the pivot value $k_{0}=0.05$ /Mpc.  

\begin{figure}[t]
\centering{
\epsfig{file=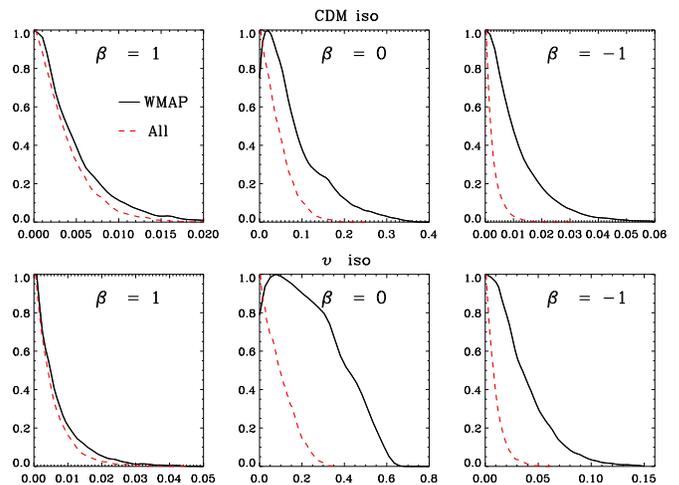,angle=90,height=6.5cm,width = 9cm}}
\caption{1-dimensional likelihood distributions for CDM (top) and neutrino (bottom) density isocurvature models 
with perfectly correlated $\beta = 1$, anti-correlated $\beta = -1$ and uncorrelated  $\beta =0$ isocurvature 
perturbations using WMAP only data (black full) and combined with SDSS, SNLS and BBN data (red dashed).
\label{oned_single}}
\end{figure}

The angular power spectrum of a given correlation 
$ij$ is given by 
\ba
C_{\ell}^{ij}=\int_0^\infty \frac{dk}{k} \Delta_{ij}^2(k)
\Theta_{\ell}^i(k)\Theta_{\ell}^j(k),
\ea 
where $\Theta_{\ell}^i(k)$ is the 
photon transfer function for initial condition $i$. 
An admixture of the adiabatic mode with a single isocurvature mode can be 
expressed in terms of the pure adiabatic, isocurvature and wholly 
correlated spectra, such that
\ba
C_\ell&=& {\cal A}_{AA}C_\ell^{AA} + {\cal A}_{II}C_\ell^{II} +2{\cal A}_{AI}C_\ell^{AI},
\ea
where $I$ labels $C,N$ or $V$. This is commonly parameterized using 
one of the following:
\ba
C_\ell&=& C_\ell^{AA} + B^{2}C_\ell^{II} +2B\cos \theta  C_\ell^{AI},\\
&=& (1-\alpha)C_\ell^{AA} + \alpha C_\ell^{II} 
+2\beta \sqrt{\alpha(1-\alpha)}C_\ell^{AI},
\label{eqn_cl}
\ea
to within an overall normalization factor, with the former used 
by \cite{Amendola:2001ni,Valiviita:2003ty,Peiris:2003ff} and the latter 
by \cite{crotty:03,beltran:04,Beltran:2005gr}. 
Using equation (\ref{eqn_cl}), the amplitude and correlation phase 
of the isocurvature contribution are given by $\alpha$ 
and $\beta$ respectively. These in turn can be related 
to the overall ratio of isocurvature to adiabatic component given by
$B={\cal S}/{\cal  R}$, with 
$\alpha= B^{2}/(1+B^{2})$, and general correlation $\beta = \cos \theta$.
We will use $\alpha$ and $\beta$ as parameters in this analysis, 
for models with a single isocurvature mode, {sampling 
with $\beta$ directly 
rather than $2\beta \sqrt{\alpha(1-\alpha)}$.
With these definitions, a positive correlation 
between the adiabatic and isocurvature perturbations will produce a
wholly correlated CMB spectrum with a negative amplitude at large 
scales, in agreement with e.g. 
\cite{Amendola:2001ni,Peiris:2003ff,Valiviita:2003ty,beltran:04} and 
the CAMB package. The primordial adiabatic perturbation may be defined 
such that these correlated spectra have the opposite sign, as used in
\cite{Bucher:2004an,Moodley:2004nz,Kurki-Suonio:2004mn,Dunkley:2005va}.

The above parameterizations do not naturally extend to 
the addition of more than one isocurvature mode, for which a 
method is given in \cite{Bucher:2004an,Moodley:2004nz}. Here  
\ba 
C_\ell = \sum_{ij=1}^N {z_{ij}} \hat{C}_\ell^{ij}
\label{z_param}
\ea
for $N$ perturbation 
modes, where $\sum_{i,j=1}^N z_{ij}^2=1$, and any 
matrix $z$ with a negative eigenvalue is assigned 
a zero prior probability.
$\hat{C}_\ell$ are normalized to have 
equal CMB power in each mode $i$, such that 
$z_{ij}$ quantify the physically observable 
power in each mode$^1$ 
\footnote{$\hat{C}_\ell^{ij}=C_\ell^{ij}/\sqrt{P_iP_j}$, 
with the power $P_i=\sum _{\ell =2}^{1000}(2\ell +1)~C^{ii}_\ell(TT)$}.
By definition the coordinates $z_{ij}$ lie on the surface of 
a $N^2$-dimensional unit sphere, which can be sampled 
uniformly with $N^2-1$ amplitude parameters using the volume 
preserving mapping shown in \cite{Moodley:2004nz}. 
To quantify the isocurvature contribution in the case of multiple modes, 
a measure is given by $r_{iso}= {z_{iso}}/{(z_{iso}+z_{AA})}$,
where $z_{iso} = \sqrt{1-z_{AA}^2}$. This is equivalent 
to the $f_{iso}$ parameter in \cite{Bucher:2004an,Moodley:2004nz,Dunkley:2005va}, but 
different to the $f_{iso}$ used in e.g. 
\cite{Peiris:2003ff,Valiviita:2003ty,Beltran:2005gr}, which 
corresponds to $B$ in equation ~(\ref{eqn_cl}).

It is important to note that constraints on $\alpha$ can 
depend strongly  on the pivot scale, whereas  the ${z_{ij}}$ parameters are independent of the pivot. As such, the $z_{ij}$ parameterization is a useful direct measure of isocurvature, whereas $\alpha$ can be misleading when the isocurvature and adiabatic spectral 
tilts are allowed to vary independently, as will be discussed in 
section \ref{gensingiso}. For ease of comparison with previous analyses, 
however,  for single components we present constraints 
for $\{\alpha,\beta\}$ (pivoted at $k=0.05/$Mpc) and give 
equivalent constraints on $\{z_{ij}\}$, since the relation between 
the single mode parameterization 
$\{\alpha,\beta\}$ and the $\{z_{ij}\}$ is 
not trivial. For multiple modes we solely present results 
using the $\{z_{ij}\}$ parameterization.  

We  parameterize our cosmological model in terms of a $\Lambda$CDM scenario using the following parameters:
$\Omega_bh^2, \Omega_ch^2, \tau, \Omega_\Lambda, b_{SDSS}$, and 
an overall scalar amplitude parameter, 
limiting our search to 
flat models with scalar fluctuations and assume 3.04 massless neutrinos species with zero chemical potential. We do not investigate here broader parameter spaces including tensor modes, running in the scalar spectral index, spatial curvature or evolving dark energy. Throughout this paper we use a single parameter
$n_s$ for both the isocurvature and adiabatic spectral index. However, we do relax this constraint in section \ref{gensingiso} where we discuss the implications of a single CDM isocurvature mode generally correlated with the adiabatic one. 

We find constraints using the 3-year WMAP CMB data alone 
\cite{Hinshaw:2006ia,Page:2006hz} for CDM and neutrino density 
isocurvature scenarios 
with a variety of degrees of correlation, and constraints
on one, two and three isocurvature components more generally  in combination with data 
from the SDSS galaxy survey \cite{Tegmark:2003uf}, including 
a Gaussian prior on the 
SDSS bias measurement $1.03\pm 0.15$ \cite{Seljak:2004sj} and non-linear corrections 
\cite{halofit}, and supernova data from SNLS \cite{Astier:2005qq}. We 
include BBN estimates of the baryon to photon ratio,
conservatively encompassing measurements of both Deuterium 
($\eta_{10}=6.4\pm0.7$) and Helium-3 ($\eta_{10}=6.0\pm1.7$) 
\cite{Steigman:2005uz} by imposing a Gaussian prior of $\Omega_bh^2=0.022\pm0.006$. 
Small-scale polarization data do not yet noticeably tighten the constraints so we do not include them in the analysis. As we discuss in the analysis and conclusion, however, future small scale polarization data could well be an important test of isocurvature scenarios.
We generate CMB and matter power spectra using the CAMB package \cite{camb}, which is consistent with our perturbation definitions.
The likelihood surfaces are explored using Markov Chain Monte Carlo methods, 
applying the spectral convergence test described in \cite{dunkley}, and 
the Gelman and Rubin convergence test \cite{gelman}. 
We also use a downhill simplex method to find the 
best-fit  likelihood, starting from the maximum likelihood point sampled by the chains, since the likelihood peak is only sparsely sampled by MCMC 
in high dimensional spaces.

\section{Uncorrelated and perfectly correlated isocurvature: single modes}\label{puresingiso}

In a variety of theoretical scenarios the isocurvature and adiabatic fluctuations in matter can arise out of 
a single mechanism and subsequently have well-defined degrees of correlation.
We first consider constraints on purely correlated ($\beta = 1$), 
uncorrelated ($\beta = 0)$ and anti-correlated ($\beta=-1$)
CDM and neutrino density isocurvature fluctuations from 
WMAP data alone (TT+TE+EE), and in combination with other datasets. 
Figure \ref{oned_single} shows that for both CDM and 
neutrino density isocurvature, purely correlated scenarios 
are the most tightly constrained by the data. 
For CDM modes with WMAP+SDSS+SNLS+BBN (and WMAP only)  we find $\alpha < 0.009 (<0.01)$ }at 95\% confidence limit (c.l.) for
purely correlated
and $<0.009 (<0.045)$ for anti-correlated modes (consistent with \cite{Seljak:2006}, 
who find $\alpha<0.005$ including Ly-$\alpha$ data). 
The corresponding limits for  neutrino density modes are $\alpha < 0.017 (<0.025)$ 
and $<0.026 (<0.083)$. The wholly uncorrelated isocurvature modes are allowed to contribute a 
much more significant fraction of the overall power, with $\alpha<0.11 (<0.26)$ and $\alpha<0.21 (<0.55)$
(95\% c.l.) for CDM and neutrino density isocurvature modes. 
We consider the statistical support for presence of isocurvature flucutations by using the 
bestfit likelihood , ${\cal L} $, calculated by comparing the theoretical spectrum predicted by the cosmological scenario, $\{C_{l}^{th}\}$, to the observed temperature and polarization maps or spectra, {\bf x}, in light of the statistical and systematic errors encoded in the covariance matrix {\bf C},
\begin{eqnarray}
{\cal L} ({\bf x}|C_{l}^{th}) =\prod_{datasets} \frac{ \exp[ {\bf x C}^{-1}{\bf x}/2]}{\sqrt{\det C}}.
\end{eqnarray}

The 3-year WMAP data is entirely consistent with no 
isocurvature contribution being required, having no improvement over the bestfit likelihood for the adiabatic scenario 
 $-2\ln {\cal L}$ = 11252 ( arising from the joint pixel/ spectrum based likelihood approach outlined in \cite{Hinshaw:2006ia,Spergel:2006hy} with total $\chi^{2} = \sum{\bf x C}^{-1}{\bf x}=3279$ for 3244 degrees of freedom and $\sum \ln \det C = 7973$ ).

For WMAP data alone significant cosmological parameter degeneracies exist between $\alpha$ and $\Omega_c h^2$ and $n_{s}$, arising because the principal effects of the isocurvature modes are modifications to the large scale temperature fluctuations.  We find no significant degeneracy between the isocurvature fraction and the optical depth to reionization.
  The impact of these degeneracies are most significant for the uncorrelated and anti-correlated CDM and 
neutrino density modes while they have only a nominal effect on the correlated
mode constraints. For the uncorrelated and anti-correlated modes $\Omega_ch^2$ is decreased by roughly 1 $\sigma$, and $n_{s}$ is increased by 2 $\sigma$ from the adiabatic value.  The addition of SDSS, SN1a and BBN datasets tighten the 
constraints by truncating these degeneracies and bringing the values back towards the fiducial values.
 The correlated CDM and nuetrino modes are little improved by the inclusion of SDSS +SNLS data because in these scenarios there is no significant degeneracy between $\alpha$ and $n_{s}$ and $\Omega_ch^2$, the two parameters that are significantly better measured by the inclusion of the complementary datasets to the CMB.

These constraints have implications for the curvaton scenario \cite{Lyth:2002my},
which includes accelerated expansion by inflation but allows for 
primordial perturbations to be generated by the decay of a 
distinct scalar field, the curvaton.  
While no unique prescription for the generation of fluctuations in 
the curvaton scenario exists, there are a range of scenarios
where the curvaton gives rise to the cold dark matter 
isocurvature perturbations, which in general predict $n_{adi}=n_{iso}$.
  The curvaton scenario does not provide a unique 
  prescription for the generation of fluctuations, however in its simplest 
  form it predicts the existence of  cold dark matter 
isocurvature perturbations with  $n_{adi}=n_{iso}$. This is because the
 curvature and entropy perturbations are related to the 
gauge invariant Bardeen variable $\xi$
\begin{eqnarray}
{\cal S}_{C} &=& 3(\xi_{CDM}-\xi)\\
{\cal R} &=& -\xi
\end{eqnarray}
The kind and amount of isocurvature depends on when the curvaton field 
decays and CDM is created \cite{Gordon:2002gv}. Scenarios in which CDM is 
generated prior to curvaton decay have $\xi_{CDM}=0$ and the entropy and 
adiabatic fluctuations perfectly anti-correlated ($\beta=1$) yielding  
$\alpha = 0.9$ which remains ruled out at high significance. If the CDM 
is to be generated by the curvaton decay then $\beta=-1$ and the amount 
of isocurvature reflects the ratio of the curvature fluctuation after 
decay to before it, $r=\xi_{before}/\xi_{after}$ which using the sudden 
decay approximation $r\approx \rho_{curvaton}/\rho_{tot}$, 
$r = \left(1+\frac{1}{3}\sqrt{\frac{\alpha}{1-\alpha}}\right)^{-1}$ \cite{Gordon:2002gv}. 
Our analysis therefore sets limits on the curvaton decay, 
with $0.97<r<1$ (95\% c.l.) from WMAP+SDSS+SNLS, comparable to the ones 
obtained by Beltran et. al. \cite{Beltran:2005gr} ($r>0.98$ 95\% c.l.), who also included Lyman $\alpha$ constraints and used a slightly different parametrization. 
For neutrino isocurvature modes generated by density perturbations 
we find $r>0.94$ (95\% c.l.).
The most practical mechanism, however, for generating neutrino 
isocurvature perturbations is through a perturbation in the 
lepton number \cite{Lyth:2002my},and subsequent non-zero 
chemical potential, not analyzed here.

\section{Generally correlated isocurvature: single modes}\label{gensingiso}
\begin{table}
\centering
\begin{tabular}{lccccc}
\hline\hline
& \mbox{\fns CI} & \mbox{\fns CI} &\mbox{\fns NID} & \mbox{\fns NIV} \\
& $n_{adi}=n_{iso}$ & $n_{adi} \ne n_{iso}$ & $n_{adi}=n_{iso}$ & $n_{adi}=n_{iso}$ \\
\hline
$r_{\mbox{\fnsc iso}}$  &  $<0.13$ & $<0.15$ &  $<0.08$ & $<0.14$ \\ 
$z_{AA}$  & $>0.989$ & $>0.999$ & $>0.996$ & $>0.987$ \\
$z_{II}$  & $<0.09$ & $<0.09$ & $<0.06$ & $<0.12$ \\
$z_{AI}$  & $0.06\pm0.07$ & $0.03\pm0.02$ & $0.0\pm0.03$ & $0.02\pm0.05$ \\
\\
${\alpha}$ & $<0.15$ & $<0.73\dagger $ & $<0.18$ & $<0.26$  \\
${\beta}$ & $0.2\pm0.3$ &$0.08\pm0.27$& $0.0\pm0.3$ & $0.08\pm0.3$  \\ \\
Added dof & $1$ & $2$ & $1$ & $1$ \\ 
$-2\ln {\cal L}$ & $11383$ & $11379$ & $11383 $ & $11381 $ \\
$\Delta(-2\ln {\cal L})$&  $ 0$ & $-4$ & $0$ & $-2$ \\
\hline
\end{tabular}
\caption{95\% upper (or lower) 
limits, or means and 68\% confidence levels, for mode 
contributions for models with generally 
correlated isocurvature. 
Scenarios using WMAP, SDSS and SNLS data  in 
which the adiabatic and isocurvature spectral indexes are both fixed to 
be identical and where they are allowed to differ are shown. 
The best-fit likelihood and the number of degrees of freedom (dof) added to  
the standard adiabatic model are shown. $\dagger$ Constraints on 
$\alpha$ are strongly dependent on the chosen pivot point in this case, 
whereas $r_{iso}$ is pivot independent, 
and therefore provides a good measure of the isocurvature contribution. 
The $\alpha$ limit is shown here solely for completeness. 
}
\label{table1} 
\end{table}

We next consider constraints on generally correlated isocurvature 
fluctuations from combined WMAP, SDSS, SN1a and BBN data, including the CDM density, neutrino density and neutrino velocity modes individually.  

\begin{figure} 
\epsfig{file=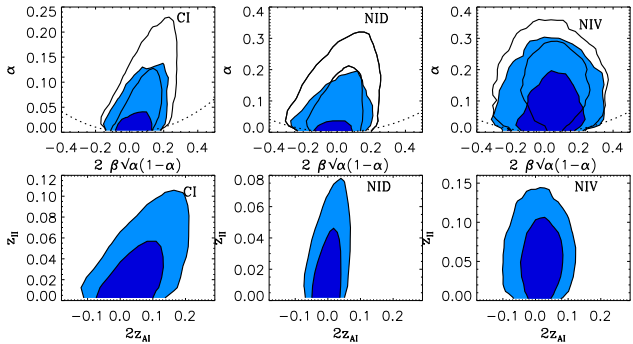,width=9cm,height=5cm}
\caption{68 \% and 95 \% 2-dimensional constraints on the amplitudes of 
generally correlated isocurvature modes, for the CI mode (left), the 
NID mode (center), and the NIV mode (right) for WMAP plus SDSS and SNLS data. 
The top panels show primordial amplitude 
contributions in terms of {$\alpha,\beta$}, using flat priors on ${z_{ij}}$ (line contours) and $\alpha,\beta$ (filled contours). The \{$\alpha, 2\beta\sqrt{\alpha(1-\alpha)}$\} parameter space is contained within a circle of unit radius shown by the dashed line.
The 
lower panels show the observable CMB power contributions in terms of ${z_{ij}}$.
\label{twod_fracs}}
\end{figure}

Constraints on the two-dimensional isocurvature amplitude and correlation 
spaces are shown in Fig. \ref{twod_fracs} and summarized in  
Table \ref{table1}. 
For CDM isocurvature we find 
$\alpha < 0.15$  at the 95\% c.l. and no overall improvement 
in the goodness of fit  $-2\ln {\cal L}$  = 11383. Neutrino density models
have $\alpha<0.18$, and neutrino velocity models $\alpha<0.26$. 
The CDM  mode prefers a small 
positive correlation with the adiabatic mode.

Repeating the analysis with the $z_{ij}$ parameters we find 95\% 
upper limits on the isocurvature fraction in terms of CMB power,  
$r_{iso}$, of 0.13 (CI), 0.08 (NID) and 0.14 (NIV) 
compared to 0.23, 0.13 and 0.24 for the first year WMAP
data \cite{Moodley:2004nz}. These constraints from 3 years of WMAP therefore show a marked improvement, being $\sim 60 \%$ of those obtained with 
the first year WMAP data; the improved 
polarization data prefer a lower level of isocurvature.

\begin{figure}
\epsfig{file=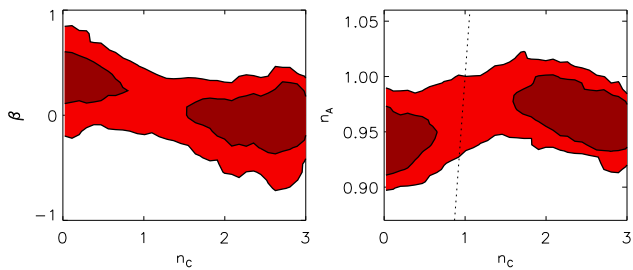,width=8.8cm,height=4cm}
\caption{The effect of varying the CDM isocurvature spectral index independently for an AD+CI scenario with WMAP plus SDSS and 
SNLS data: 
68 \% and 95 \% constraints on the AD-CI cross-correlation 
$\beta$ (left panel), and adiabatic spectral index (right panel). 
The dotted line in the right panel shows $n_{adi}=n_{iso}$.
\label{niso}}
\end{figure} 

The data is fully consistent with 
$\alpha=0 / r_{iso} = 0$, with the goodness of fit   $-2\ln {\cal L}$ 
 improved by only 
$\sim 1$ for each case. These additional degrees of freedom, however, 
cause the baryon density and spectral index mean values to
move more than $1 \sigma$ from their adiabatic 
values: both values are increased by $2 \sigma$ when the NIV mode is included,
exploiting the degeneracy observed in \cite{Bucher:2004an}. 

\begin{figure}
\epsfig{file=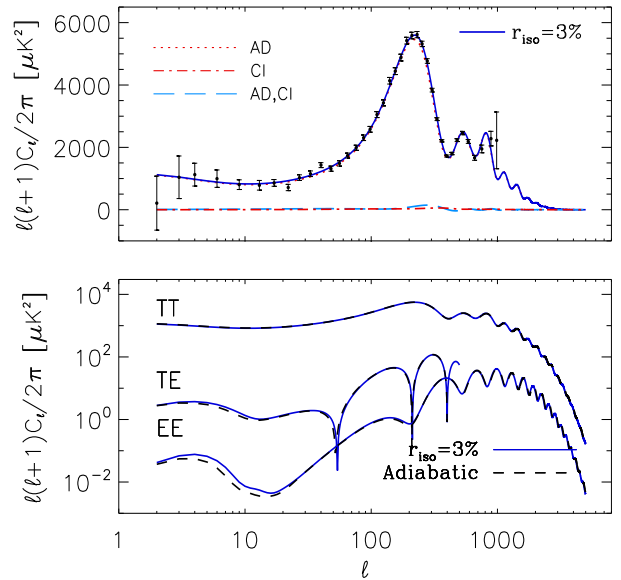,width=9cm,height=8cm}
\caption{{\it AD+CI:} (Top) CMB temperature power spectrum 
for the best-fit model with correlated AD+CI modes and independent spectral 
indices (solid) for WMAP plus SDSS and SNLS data. The model has 
$r_{iso}=3.4\%$, 
$\alpha=0.49$, $\beta=-0.26$, 
$n_{adi}=0.97$, $n_{iso}=2.7$,
and has  $-2\ln {\cal L}$ lower than the best-fit adiabatic model. 
The contribution of each mode correlation to the total spectrum is shown, 
including the WMAP data. 
Bottom: the CMB spectra are compared to the pure adiabatic 
best-fit model (dashed). Only the $\ell<500$ section of the 
TE spectrum is shown. 
\label{cdm_niso_spectra}}
\end{figure}

Our investigation finds that the results are sensitive to the choice 
of prior: constraints on $\alpha$ obtained by 
sampling $\alpha$ and $\beta$ directly differ from those derived 
from the distribution sampled using the $z_{ij}$ parameterization.
This is demonstrated in the top row of Figure \ref{twod_fracs}, where the two 
methods are compared. We see that there is more phase space available for 
models with larger $\alpha$ when sampling with a uniform prior 
on the observable isocurvature CMB power, than there is when sampling 
with a uniform prior on $\alpha$. The likelihood of the best-fitting 
models are not affected by the choice of prior however, and we can expect the 
dependence to be reduced as data improves.

If the assumption of $n_{iso}=n_{adi}$ is relaxed, and $n_{iso}$ is allowed to vary freely within the bounds $0<n_{iso}<3$, a large phase space is opened up for 
models with large $n_{iso}$, within the range allowed by the prior. The 1-D marginalized constraints on the isocurvature contribution are increased from $r_{iso} <0.13$ to $<0.15$ (95\% c.l.). The isocurvature tilt cannot be constrained by WMAP+SDSS+SN1a datasets alone, however \cite{Beltran:2005gr} show that additional Ly-$\alpha$ data prefer higher tilts of $1.9\pm1$. When spectral indices are able to vary freely, $r_{iso}$ is a good measure of isocurvature because $\alpha$ then becomes extremely sensitive to the pivot point at which the spectral indices are defined. The scenario 
we investigate here with CDM isocurvature and $n_{iso}\neq n_{adi}$ is a case in point. For models with high $n_{iso}(0.05/Mpc)$ the isocurvature power on larger cosmological scales is significantly reduced for a given $\alpha$ and therefore $\alpha$ is able to be increased to compensate. The relative power in isocurvature, however, roughly indicated by $\sim\alpha/(1-\alpha)\times (k/k_{0})^{n_{iso}-n_{adi}}$, 
is not increased.

The best-fit model is shown in  Figure \ref{cdm_niso_spectra}, has $n_{adi}=0.97$, $n_{iso}=2.7$ and may be distinguished from the adiabatic 
model at large scales in polarization.
The goodness of fit is improved, compared to the adiabatic model, 
by a $\Delta(-2\ln {\cal L})$ 
of 4 to $11379$ for 2 additional degrees of freedom. 
Such an improvement is driven by the use of the 
3-year WMAP data in the analysis. 
Future measurements will help 
distinguish these, currently degenerate, high isocurvature tilt models.

\begin{figure}
\hskip +0.1in
\epsfig{file=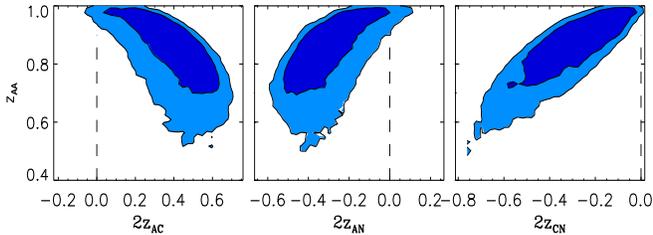,width=9cm,height=4cm}
\caption{ {\it AD+CI+NID:} 2-dimensional constraints show the 
degeneracy between isocurvature cross-correlation amplitudes  $z_{ij}$, 
and the adiabatic amplitude $z_{AA}$, that allows the destructive 
interference of isocurvature spectra.}
\label{twod_iso_double}
\end{figure}

This subset of models may inform us about double 
inflation scenarios. If there are multiple fields driving 
inflation, it is possible to generate entropy perturbations in which the 
additional light fields modify the curvature perturbations on horizon 
scales, and also modify the consistency relations relating scalar to 
tensor modes \cite{Bartolo:2001rt,Wands:2002bn}.
In the most general case $n_{adi} \ne n_{iso}\ne n_{corr}$, with each related to the slow roll parameters along the flat directions of the scalar fields. 
For example, for theories in which one scalar field plays a dominant role in 
driving inflation, but two fields play a significant role during 
reheating, one finds that $n_{iso}\approx n_{corr}$ \cite{Wands:2002bn}. 
Analyses in which $n_{corr}$ has been allowed to vary have found it is a 
nuisance parameter unconstrained by data \cite{beltran:04}, so 
we believe it is reasonable to assume that fixing the scale 
dependence of the cross 
spectrum , $n_{corr}=(n_{adi}+n_{iso})/2$, does not unduly bias the conclusion.
For two field inflation of two minimally coupled scalar fields of 
mass $m_{h}$ and $m_{l}$, the magnitude and correlation of the 
resulting CDM isocurvature component are dependent on the ratio of the 
masses  $R = m_{h}/m_{l}$  and number of e-foldings $s_{k}$. A bound on 
$R$ comes from the magnitude of the cross-correlated spectrum \cite{beltran:04}
\begin{eqnarray}
2\beta\sqrt{\alpha(1-\alpha)}|_{max} &=& \frac{2s_{k}(R^{2}-1)}{s_{k}^{2}+(R^{2}+1)^{2}}
\end{eqnarray} 
 Assuming $s_{k}=60$ we find an upper bound on the ratio of the 
two scalar fields of $R< 3.5$. This is weaker than the $R<3$ at 
95\% c.l. obtained 
with the inclusion of Lyman-$\alpha$ data \cite{Beltran:2005gr}, 
although we caution that constraints 
on $\alpha$ in this case are strongly dependent on the choice of 
prior and pivot scale.

It would be interesting, but beyond the scope of this paper, to place constraints on specific double inflation models in which model-dependent predictions for each mode's spectral index are included. 
\begin{figure}[t]
\hskip +0.1in
\epsfig{file=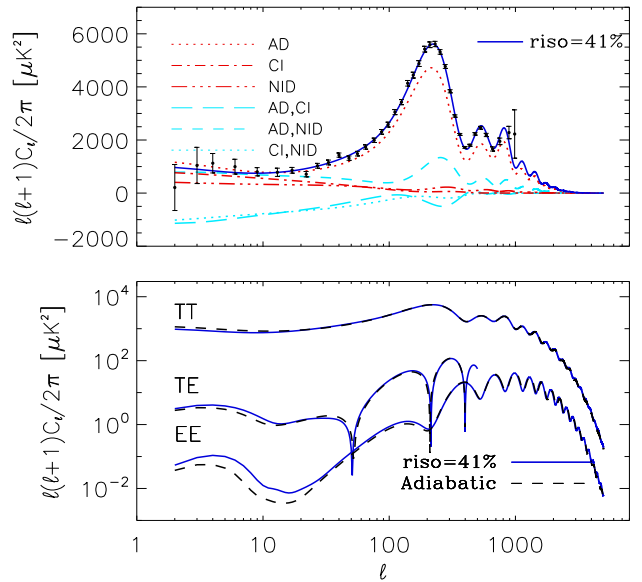,width=9cm,height=8cm}
\caption{{\it AD+CI+NID:} Spectra as in Figure \ref{cdm_niso_spectra}. 
The model shown has correlated modes, with $r_{iso}=41\%$, and
cosmological parameters 
$\Omega_bh^2=0.026$, $\Omega_ch^2=0.12$, $\Omega_\Lambda=0.73$,
$\tau=0.10$, $n_s=0.92$, $b_{SDSS}=0.98$.  
The isocurvature spectra add destructively, canceling almost completely.
\label{cinid_spectra}}
\end{figure}

\section{Generally correlated isocurvature: multiple modes}\label{mixed}

In this section we consider models with two additional correlated 
isocurvature modes, (CI+NID, CI+NIV, NID+NIV), and finally a model with 
the full set of adiabatic and three isocurvature modes.  
We sample the modes using the parameterization 
given in equation (\ref{z_param}) for $N=3$ and $N=4$.
Table \ref{table2} shows constraints for the relative mode 
contributions for this set of models. 
We also give the primordial amplitudes ${\cal A}_{ii}$ of the 
auto-correlations contributing to the best-fitting models, where 
$C_\ell= \sum_{i,j=1}^{N} {\cal A}_{ij}C_\ell^{ij}$.

\begin{figure}[t]
\hskip +0.1in
\epsfig{file=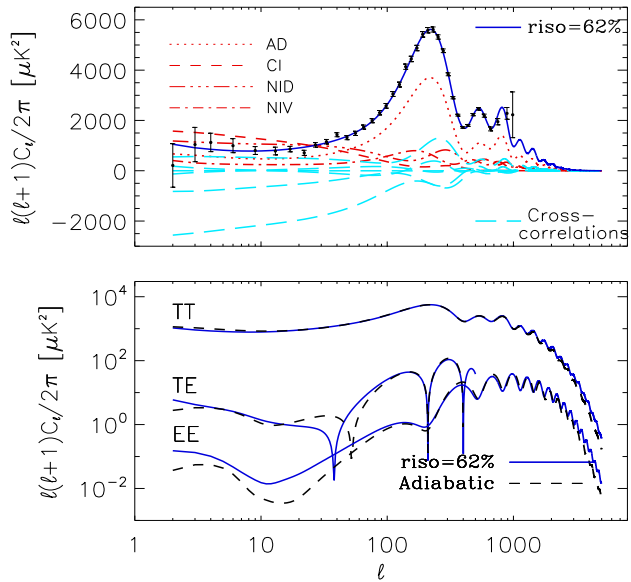,width=9cm,height=8cm}
\caption{{\it AD+CI+NID+NIV:} Spectra as in Figure \ref{cdm_niso_spectra}. 
The model shown has $r_{iso}=62\%$. No BBN or bias constraint has 
been imposed; the cosmological parameters are
$\Omega_bh^2=0.037$, $\Omega_ch^2=0.13$, $\Omega_\Lambda=0.75$,
$\tau=0.10$, $n_s=0.98$, $b_{SDSS}=1.2$. 
\label{all_spectra}}
\end{figure}

{\it Two isocurvature modes:} As is shown in Table  \ref{table2}, models 
with two modes permit far more isocurvature than those with a single 
mode.

Although when considered individually, the neutrino velocity isocurvature 
modes allow the largest isocurvature fraction, interestingly when two modes 
are included, joint CDM and neutrino density 
isocurvature (CI+NID) allow the most freedom, more than twice  
as much isocurvature ($r_{iso}=0.4\pm0.1$) as the combinations 
including the neutrino velocity mode. This freedom arises from 
degeneracies within the isocurvature components destructively 
interfering, originally observed in \cite{Moodley:2004nz} and 
shown in Figure \ref{twod_iso_double}.  Degeneracies with the NIV mode do also 
exist where the spectra add constructively, but such models have 
large baryon densities ruled out by current BBN measurements.


Figure \ref{cinid_spectra} shows the best-fit CMB temperature spectrum for the CDM + neutrino density isocurvature  model with $r_{iso}=41\%$, compared with 47\% with WMAP year 1 data  \cite{Moodley:2004nz}. The contributions from all six correlations 
(three auto-correlations and three cross-correlations) lead to greater large scale polarization CMB power, but 
cancel almost completely in the CMB temperature and galaxy power spectrum.

All three two-mode models prefer 
baryon densities higher than the concordance value (with mean values 
$0.025 < \Omega_bh^2 < 0.027$), despite the BBN constraint. The spectral index is also more poorly
constrained, with the CI+NID models preferring 
a low spectral index ($0.93\pm0.03$), and the CI+NIV ($0.99\pm0.04$) and 
NID+NIV ($1.03\pm0.03$) preferring larger values. The other cosmological parameters are consistent with adiabatic $\Lambda$ CDM.

{\it Three isocurvature modes:} When all three modes are included, the constraints are highly sensitive to the BBN constraint due to the strong degeneracy between $\Omega_bh^2$
and the NIV amplitude. With the BBN prior and SDSS bias we find 
$r_{iso}=0.44\pm0.09$, increasing to $r_{iso}=0.51\pm0.09$ when no BBN or 
SDSS bias priors are included, to be compared to $0.57\pm0.09$ found for the first year WMAP data  \cite{Bucher:2004an}. 

A well-fitting model ($\Delta$($-2\ln{\cal L}$)=-6 with 10 extra degrees of freedom 
in comparison to the adiabatic best fit) with a majority of the 
power coming from isocurvature modes, $r_{iso}=62\%$, 
is shown in Fig. \ref{all_spectra}.
This model was obtained without the BBN prior and 
has a large baryon density ($\Omega_bh^2 =0.037$); however even including the 
BBN constraint,  the baryon density is higher than the concordance value 
for this class of models ($0.031\pm0.003$), and $\Omega_ch^2$ 
is raised by $1\sigma$ to $0.124\pm0.007$. 
Figures \ref{cinid_spectra} and 
\ref{all_spectra} indicate that precision small-scale temperature and 
large-scale polarization will to more tightly determine the 
underlying initial conditions. In particular future small-scale 
CMB experiments 
should strengthen the constraint on the baryon density, since 
high baryonic isocurvature models are more degenerate at 
smaller scales than their low baryon counterparts. 

Given that the data can still only poorly constrain the isocurvature contribution 
for these multi-mode models, the constraints presented here depend on the 
prior distribution we have chosen. Previous work has shown that other 
parameterizations can decrease or increase this contribution \cite{Moodley:2004nz}, depending on the phase-space volume available for purely adiabatic models compared to mixed isocurvature models. As with the single mode case, improved data will help limit this prior dependency.

\section{Conclusion}\label{Conclusion}

\begin{table*}[t]
\centering
\begin{tabular}{lcccccc}
\hline\hline
& \mbox{\fns CI+NID} & \mbox{\fns CI+NIV} &\mbox{\fns NID+NIV} && \mbox{\fns CI+NID+NIV} & \mbox{\fns  CI+NID+NIV} \\  
&&&&&& No BBN/bias\\
\hline
$r_{\mbox{\fnsc iso}}$  &  $0.4\pm0.1$ &  $0.15\pm0.06$ & $0.20\pm0.08$ &&  $0.44\pm0.09$ & $0.51\pm0.09$\\
$z_{AA}$  & $0.8\pm0.1$ & $0.98\pm0.02$ & $0.96\pm0.03$ && $0.8\pm0.1$ & $0.7\pm0.1$\\
$z_{CC}$  & $0.2\pm0.1$ & $0.04\pm^{0.03}_{0.02}$ & $-$ && $0.21\pm0.09$ & $0.23\pm0.09$\\
$z_{NN}$ & $0.16\pm0.09$ & $-$ & $0.05\pm0.03$ && $0.23\pm0.09$  & $0.28\pm0.10$\\
$z_{VV}$ & $-$ & $0.09\pm0.05$ & $0.17\pm0.09$ && $0.15\pm0.06$  & $0.21\pm0.09$\\
 \\
$10^{10}{\cal A}_{AA}$ & $18.1$ & $21.8$ & $20.0$ && $20.0$ & $14.1$\\
$10^{10}{\cal A}_{CC}$ & $5.2$ & $0.45$ & $-$ && $11.2$ & $12.4$\\
$10^{10}{\cal A}_{NN}$ & $8.1$ & $-$ & $0.35$ && $34.9$ & $37.2$\\
$10^{10}{\cal A}_{VV}$ &  $-$ & $3.0$ & $4.7$ &&  $8.9$ & $21.4$\\ \\
Added dof & $4$ & $4$ & $4$ && $8$ & $10$ \\
 $-2\ln {\cal L}$ ($\Delta(-2\ln {\cal L})$)
 & $11379 (-4)$ & $11382 (-1)$ & $11381 (-2)$ && $11375 (-8)$  & $11374 (-9)$\\
\hline
\end{tabular}
\caption{Means and $68\%$ c.l. 
for auto-correlated contributions ($z_{ij}$) for models with generally 
correlated isocurvature modes as indicated. The best-fit primordial 
amplitudes ${\cal A}_{ii}$ for the pure modes are shown. 
\label{table2}}
\end{table*}

We have investigated the constraints on the presence of a variety of 
isocurvature modes in the initial conditions of structure formation, in light of recent observations of 
temperature and polarization CMB data and large scale structure and supernovae surveys.

The improved WMAP data, with the inclusion of low $l$ polarization measurements, has strengthened these constraints on the contributions of individual isocurvature modes, with the polarization data disfavoring models with a large isocurvature fraction. Scenarios with either CDM (or baryon) or neutrino isocurvature 
allow only a very limited contribution, which can be translated into strong constraints on the curvaton model and some double-field inflationary models. 
Although  models with multiple isocurvature modes do not offer a significantly better fit to the data,  models with non-zero isocurvature fluctuations fit the data as well as the adiabatic model and can comprise the majority of power when additional modes are considered 
simultaneously. 

Of the models with more than one isocurvature mode, those most likely to pose the greatest difficulty for 
distinguishing with future data are those
with large fractions of both correlated 
CDM and neutrino density isocurvature, which provide the best fit to 
the data, and due to 
their destructive interference are highly degenerate in the CMB and galaxy 
power spectra. Those with 
neutrino velocity fluctuations (both two and three mode models) 
are better constrained by BBN and bias measurements.
 
With WMAP plus LSS and SN data, the baryon density and spectral tilt are 
found to be sensitive to the inclusion of isocurvature modes. With 
the current data, the reionization optical depth however is robust despite the modifications isocurvature models can make to large scale polarization spectra. Extending beyond the $\Lambda$ CDM scenario, given results found in \cite{dunkley}, we would not 
expect that allowing independent tilts for 
all the modes, or including curved geometries, 
to have a large effect. Future B-mode polarization data will help break degeneracies between tensor and isocurvature modes that would currently arise from large scale temperature CMB data.

Future small-scale temperature and polarization data, 
together with improved galaxy and Lyman-$\alpha$ power spectrum measurements, 
should help constrain a subset of the models we have considered, 
but improved large scale 
CMB polarization data from WMAP and in particular Planck, demonstrated 
in \cite{Bucher:2002}, 
will be crucial if we are to 
strongly constrain this general set of correlated isocurvature models.

\section*{Acknowledgments}
We would like to thank Olivier Dore, Lyman Page and David Spergel for helpful 
discussions and comments in preparing this paper. We acknowledge the use of the 
CAMB code. We thank Ian Sollom, Mike Hobson, and Anthony Challinor for pointing out a numerical error in Section 4 after publication. RB is supported by NSF grant AST-0607018. JD acknowledges support from NASA grant LTSA03-0000-0090.
EP is an ADVANCE fellow (NSF grant AST-0340648), also supported by NASA grant 
NAG5-11489.


\end{document}